\documentclass[prl,aps,twocolumn,superscriptaddress,nofootinbib]{revtex4}

\usepackage{epsfig}
\usepackage{amsfonts}
\usepackage{latexsym}
\usepackage{amsmath}
\usepackage{amssymb}
\usepackage{slashed}
\usepackage{longtable}
\usepackage{mathrsfs} 
\usepackage[normalem]{ulem}

\usepackage{graphicx}
\usepackage{hyperref}

\numberwithin{equation}{section}

\newcommand{\bea}{\begin{eqnarray}}
\newcommand{\eea}{\end{eqnarray}}
\newcommand{\be}{\begin{equation}}
\newcommand{\ee}{\end{equation}}
\newcommand{\ba}{\begin{align}}
\newcommand{\ea}{\end{align}}

\def\Or[#1]{{\text{O}}\left({#1}\right)}
\def\dotl[#1,#2]{\left\langle #1, #2 \right\rangle}
\def\dotlb[#1,#2]{[ #1, #2 ]}
\def\dotp[#1,#2]{(#1) \cdot (#2)}
\def\aff[#1,#2]{\hat{#1}(#2)}
\def\n4sym{{\cal N}=4 SYM}
\def\>{\rangle}
\def\<{\langle}
\def\weight[#1,#2,#3]{\{(#1),#2,#3\}}
\def\ads[#1]{$\text{AdS}_{#1}$}

\newcommand{\vev}[1]{\left\langle{#1}\right\rangle}

  \makeatletter
  \let\over=\@@over \let\overwithdelims=\@@overwithdelims
  \let\atop=\@@atop \let\atopwithdelims=\@@atopwithdelims
  \let\above=\@@above \let\abovewithdelims=\@@abovewithdelims

\usepackage{xcolor}

  \newcommand{\cB}{{\cal B}}
  
\newcommand{\cE}{{\cal E}}

\begin{document}

\preprint{}

\title{Quantum Hall effective action for anisotropic Dirac semi-metal}

\author{Carlos Hoyos} 
\affiliation{Department  of  Physics  and  Instituto  de  Ciencias  y  Tecnolog\'{i}as  Espaciales  de  Asturias  (ICTEA)\\
Universidad  de  Oviedo,  c/  Federico  Garc\'{i}a  Lorca  18,  ES-33007  Oviedo,  Spain}
\author{Ruben Lier} 
\author{Francisco Pe\~{n}a-Benitez} 
\author{Piotr Sur\'owka} 
\affiliation{Max-Planck-Institut  f\"ur Physik komplexer Systeme, N\"othnitzer Str. 38, 01187 Dresden, Germany}

\begin{abstract}
We present a study of Hall transport in semi-Dirac critical phases. The construction is based on a covariant formulation of relativistic systems with spatial anisotropy. Geometric data together with external electromagnetic fields is used to devise an expansion procedure that leads to a low-energy effective action consistent with the discrete $PT$ symmetry that we impose. We use the action to discuss terms contributing the Hall transport and extract the coefficients. We also discuss the associated scaling symmetry.
\end{abstract}

\maketitle
\section{Introduction}

Low-energy effective actions have become an important tool to study topological responses of quantum phases of matter \cite{Hansson2020}. The predictive power of effective actions comes from the fact that they mainly rely on the symmetries of the system in question, hiding our ignorance about the microscopic details of the system in a set of parameters. Although initially applied to phases with large symmetry groups such as Galilean or Poincar\'e, in recent years effective theories have been used to shed light on states with more exotic or reduced symmetry groups. These symmetries can be realized in various topological materials undergoing quantum phase transitions from a conductor to a band insulator. They emerge when topological defects occurring at isolated points in momentum space collide. Such defects are symmetry protected points where valence and conduction bands touch.

\begin{figure}[t!]
  \begin{center}
  \includegraphics[scale=1.3]{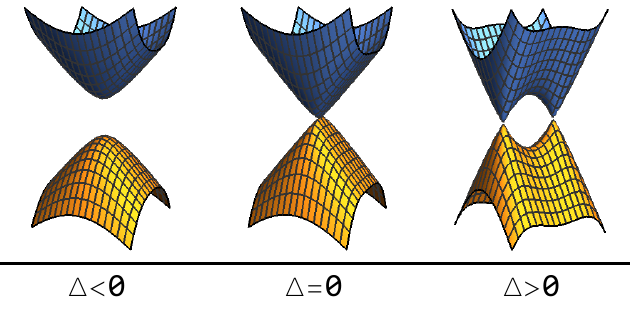}
  \caption{Spectrum of the Hamiltonian. $\Delta=0$ corresponds to the semi-Dirac critical point.}\label{fig_spectrum1}
  \end{center}
  \end{figure}

An interesting class corresponds to the quantum critical point connecting a band insulator and a graphene-like state\footnote{At the critical point, the $\pm\pi$ Berry's phase of each Dirac cone annihilate.} (see Fig. \ref{fig_spectrum1}).  The critical phase, known as the semi-Dirac phase,  is semi-metallic with electrons dispersing linearly in one direction and quadratically in the other \cite{banerjee2009pickett,depplace2010montambaux,dietl2008montambaux,Pena-Benitez:2018dar,Uryszek,Tarruell2012,Sato2019,PhysRevX.8.011049,Mawrie2019}. Examples of systems that exhibit such phases in two dimensions include  TiO$_2$/VO$_2$ heterostructures \cite{pardo2009pickett}, (BEDT-TTF)$_2 I_3$ organic salts under pressure \cite{katayama2006suzumura}, photonic metamaterials \cite{wu2014} and certain non-Hermitian systems \cite{Narayan2020}. Semi-Dirac phases reveal distinct features, for example, in transport phenomena \cite{Link:2017ora} or driven by light \cite{Narayan2015,Saha2016}.

In \cite{Pena-Benitez:2018dar} it has been shown numerically that when a magnetic field is applied to these systems, the topological response of such phases is governed by two independent non-dissipative viscosities, in contrast to isotropic phases where a single Hall viscosity is present \cite{1997physics12050Avron,avron1995zograf}. However, the method did not provide an explanation for the vanishing of a possible third viscosity, which is allowed by the continuous symmetries.

To shed light on the generic low energy properties of such quantum Hall states, we construct an effective action for systems with a semi-Dirac phase. Our construction contributes to the quest of understanding Hall viscosities; non-dissipative transport coefficients that emerge in the context of topological order \cite{avron1995zograf,read2009,Hoyos:2014pba,Hoyos:2011ez,Golan:2019svj,Rose:2020xfb,B_ttcher_2019}, fluid dynamics \cite{Alekseev2016,Delacretaz:2017yia,Lucas:2014sia,Ganeshan2017,Matthaiakakis,Narozhny2019,Stern2019,Jain2020,Rao2020,Wiegmann:2013hca,Lapa2014,Abanov2019} or active matter \cite{Banerjee2017,Salbreux2017,PhysRevE.101.052606,Vitelli2020,Banerjee2020,Markovich2020}. Initially thought as an elusive transport property, Hall viscosity has been experimentally identified in both hard- \cite{Berdyugin2019} and soft- \cite{Soni2019} condensed matter experiments.

In this paper we present a step-by-step construction of the low-energy effective theory. First we study the coupling of fermions to the background geometry and gauge fields in a microscopic model. The coupling  to the curved background is not unique, unless extra geometric constraints are invoked \cite{Copetti:2019rfp}, but this ambiguity is hidden in the coefficients of the effective action. In the following section we present the main construction of the effective action including symmetries, derivative counting and contributing terms up to the second order in the expansion. The main properties of the effective action derive from the breaking of the would-be $2+1$ Lorentz group of a graphene-like phase to a $1+1$ Lorentz subgroup by a spatial vector that marks the anisotropic direction, and by an unbroken discrete $PT$ symmetry. The technical details are relegated to the Supplement. Finally, we present the topological transport of the semi-Dirac states that follows from our effective action. We show that a hitherto overlooked transport coefficient, due to antisymmetric strains is necessarily present in such phases. We close with a discussion of the results.

\section{The semi-Dirac system} \label{sec:SD}

The low energy Hamiltonian describing the anisotropic semi-Dirac semimetal reads
\begin{equation}\label{eq_Hamiltonian}
\mathcal H = \mathbf d(\mathbf p)\cdot\boldsymbol\sigma\,,
\end{equation}
where $\boldsymbol \sigma$ is a vector of Pauli matrices. $\mathbf{d}(\mathbf{p})=(p_x,\frac{p_y^2}{2m}-\Delta,0)$ with $m$ being a parameter with dimensions of mass. In Fig. \ref{fig_spectrum1} we show the three phases captured by the Hamiltonian Eq. \eqref{eq_Hamiltonian}, where the semi-Dirac phase is obtained when $\Delta=0$.

When we study the coupling to background fields it is more convenient to use the action formalism
\begin{equation}\label{eq_action1} 
 S = \int d^3x \left( i\bar\psi\gamma^a\mathcal P_a^b\partial_b\psi-\bar\psi\,M[\Delta,l^a\partial_a]\psi\right)\,,
 \end{equation}
 where $\gamma^a=(\sigma_3,-i\sigma_2,i\sigma_1)$ are $2+1$ Dirac matrices. One can view the system as having an effective mass which is momentum dependent, i.e.  $M(\Delta,l^a\partial_a) = \Delta -\frac{1}{2m}(l^a\partial_a)^2$. We aim to have the theory in a covariant form, which will facilitate putting the electrons on a curved space-time. Therefore we have introduced the anisotropy vector $l^a=(0,0,1)$, and the transverse projector
 \begin{equation}
\mathcal P_a^b=\delta_a^b-l_al^b\,,\qquad l^al_a=1\,.
\end{equation}
$l^a$ breaks the $SO(2,1)$ Lorentz symmetry present for standard relativistic fermions down to $SO(1,1)\times \mathrm C_2$, where $SO(1,1)$ corresponds to boosts in the plane transverse to $l^a$ and $\mathrm C_2$ $180^o$ rotations on the spacelike plane containing vector $l^a$. In fact, such a discrete rotation is equivalent to the transformation $l^a\to -l^a$. The discrete symmetries can be understood from the point of view of a graphene-like model as proposed in \cite{Pena-Benitez:2018dar}. Ignoring the spin microscopic degrees of freedom, it consist of two species of fermions organized by valleys on two hexagonal sub-lattices. Such fermions exhibit parity $P$ and time reversal symmetry $T$. Anisotropy stems from the mass deformation that breaks separate symmetries, however, preserving the $PT$ combination. In addition there is the graphene's inversion symmetry that we refer to as $P^*\simeq C_2$ that acts by sending $\vec x\to-\vec x$, and simultaneously interchanging the sub-lattice atoms and valleys \cite{Gusynin:2007ix}. 

In the semi-Dirac phase $\Delta=0$ the action enjoys an additional anisotropic scaling symmetry
\be
{\cal P}_a^b x_b \to \lambda^2 {\cal P}_a^b x_b , \ \ l^ax_a \to \lambda l^a x_a,\ \ \psi \to \lambda^{-3/2}\psi.
\ee
The scaling dimensions deduced from the fermionic action for the derivatives and gauge potentials are
\be
\left[{\cal P}_a^b\partial_b\right]=\left[{\cal P}_a^b  A_b\right]=2,\ \ \ \left[l^a\partial_a\right]=[l^aA_a]=1.
\ee
Then, for $l^a=(0,0,1)$ the field strenghts scaling dimensions are $[E_y]=[B]=3$ and $[E_x]=4$, while the energy has dimensions $[E]=[\partial_t]=2$.  The scale invariance is reflected  on the dependence of the energy of Landau levels with the magnetic field $E\sim B^{2/3}$  \cite{halperin2009esaki}.

\section{Coupling to background fields} \label{sec:riem}

As already pointed out, our goal is to construct the generating functional of the semi-Dirac system in the presence of magnetic field. To do so, it is necessary to couple the fermionic action Eq. \eqref{eq_action1} to external $U(1)$ gauge fields and a curved background which will allow us to predict the response of the $n-$point functions of the $U(1)$ and the stress energy tensor by functional differentiation.

In the first place we will introduce a background metric $g_{\mu\nu}$ and an orthonormal basis $e_a$ of tangent vectors, $g(e_a,e_b)=\eta_{ab}$ (inverse vielbeins). In addition we have introduced a set of dual cotangent one forms $E^a$ (vielbeins), such that 
\begin{equation}
\langle E^a,e_b\rangle =\delta^a_b\,.
\end{equation}
The metric can be determined by the vielbeins $g_{\mu\nu}=\eta_{ab} E_{\ \mu}^a E_{\ \nu}^b$. Then, the generalization of Eq. \eqref{eq_action1} to curved spacetime involves replacing the derivative, Dirac matrices and vector $l^a$ in flat space by their pullbacks to the tangent space $\partial_a=e_a\,^\mu\partial_\mu$, etc. 

Actually, given the anisotropic nature of the system it is convenient to split the vielbeins in their transverse and longitudinal part
 \begin{eqnarray}
 \hat e_a\,^\mu &=&\mathcal  P_a^b e_b\,^\mu\,, \qquad\qquad \hat E^a\,_\mu =\mathcal P^a_b E^b\,_\mu\,,\\
 l^\mu &=& l^a e_a\,^\mu\,,\qquad\qquad l_\mu = l_a E^a\,_\mu\,.
 \end{eqnarray}

It is also convenient to decompose the $2+1$ Poincar\'e generators $(P_a,J_{ab})$ in terms of the broken and unbroken ones, therefore we introduce the following spliting
\begin{eqnarray}
\hat K &=& \frac{1}{2}\hat \epsilon^{ab}J_{ab}\,,\quad \hat \jmath_a = J_{ab}l^b\\
\pi &=& P_a l^a\,, \quad\quad \hat p_a = P_b\mathcal P_a^b\,,
\end{eqnarray}
where the transverse fully antisymmetric tensor is defined as $\hat\epsilon^{ab} =\epsilon^{abc}l_c$\footnote{We define $\epsilon^{abc}$ as the fully antisymmetric symbol, therefore $\varepsilon^{\mu\nu\rho}=\epsilon^{abc}e_a\,^\mu e_b\,^\nu e_c\,^\rho$ has components $\sqrt{-g} \varepsilon^{012}=1$}. $\pi,p_a$ generate longitudinal and transverse translations respectively and $\hat K$ is the $SO(1,1)$ boost generator. The broken generators have been collected in $\hat \jmath_a$.

The standard way of minimally coupling relativistic fermions to gauge fields and curved space-time relies on replacing the partial derivative by a covariant one
\begin{equation}\label{eq_covderrel}
\partial_\mu \to \mathcal D_\mu\equiv \partial_\mu + i e A_\mu + \frac{1}{2}\omega^{ab}\,_{\mu}J_{ab}\,,
\end{equation}
where $A_\mu$ is the $U(1)$ gauge field, $J^{ab}=\frac{1}{4}[\gamma^a,\gamma^b]$ the Lorentz algebra generators, and $\omega^{ab}\,_\mu$ the so-called Levi-Civita spin connection, defined as
\begin{equation}
\omega^{ab}\,_{\mu}=- E^a\,_\nu\nabla_\mu e^{b\nu} \,,
\end{equation}
where we set the torsion to be vanishing and choose $\nabla_\mu$ to be a covariant derivative constructed with the Christoffel symbols. Notice that this definition is covariant under the full $SO(2,1)$ group of local Lorentz transformations, while the system is only invariant under the $SO(1,1)$ subgroup that keeps the vector $l^a$ fixed, so this is not the most general possible form of the covariant derivative. To generalize it, it is convenient to split the spin connection in its longitudinal and transverse part
\begin{equation}
\omega^{ab}\,_{\mu}=\hat\omega_\mu\hat\epsilon^{ab} + 2\hat\theta^{[a}\,_\mu l^{b]}\,,
\end{equation}
where $\hat\omega_\mu=\frac{1}{2}\hat\epsilon_{ab}\omega^{ab}\,_{\mu}$, and $\hat\theta^{a}\,_\mu=\omega^{ab}\,_{\mu}l_b$, which after plugging it in the covariant derivative produces
\begin{equation}
\mathcal D_\mu=\partial_\mu + i e A_\mu + \hat\omega_{\mu} \hat K+\hat\theta^a\,_\mu \hat \jmath_a\,.
\end{equation}
In particular, $\hat \theta^a\,_\mu$ takes the form
\begin{equation}
\hat \theta^a\,_\mu = \hat e^{a \nu}  T_{ \mu  \nu }  + l_{\mu}   \hat{e}^{a \nu} T_{\nu}  ,
\end{equation}
where $T_{\mu\nu}=\partial_\mu l_\nu- \partial_\nu l_\mu$ and $T_{\mu} = l^{\nu} T_{\nu \mu}$. Actually, when a Lorentz transformation generated by $\hat K$ is applied
\begin{equation}
\hat E^a\,_\mu \to \Omega(x)\hat\epsilon^a\,_b\hat E^b\,_\mu\,, \quad l_\mu \to  l_\mu
\end{equation}
the transverse part of the spin connection transforms as an abelian connection
\begin{equation}
\hat\omega_{\mu} \to \hat\omega_{\mu}+\partial_\mu\Omega\,
\end{equation}
whereas $\hat\theta^a\,_\mu$ transforms as a vector.

 At this point we are ready to couple the fermionic field to the background geometry, introducing a more general covariant derivative which transforms properly under $SO(1,1)$ boost and local $U(1)$ gauge transformations,
\begin{equation}
 \mathcal D_\mu=\partial_\mu -i A_\mu + \frac{1}{2}\omega_\mu\,^{ab} J_{ab}+\alpha \hat\theta_\mu\,^a \hat \jmath_a\,
\end{equation}
 where $\alpha$ is an unknown coupling constant, whose value could be determined by finding an UV completion. However, from our effective field theory perspective, its specific value is irrelevant, because it would just be hidden in the particular values of the coefficients appearing in the effective action. With this covariant derivative we can write the gauge and diffeomorphism invariant action
   \begin{equation}\label{eq_action2}
 S =\int  \sqrt{-g} \left[ i\bar\psi\gamma^a\hat e_a\,^\mu\mathcal D_\mu\psi-\Delta\bar\psi\psi+\frac{1}{2m} l^\mu l^\nu\mathcal D_\mu\bar\psi\mathcal D_\nu\psi \right]\,.
 \end{equation}
The action is invariant under local gauge, diffeomorfism and boost transformations, that act on the external fields as follows

 \begin{eqnarray}\label{eq:localtrans}
 \delta A_\mu &=& \mathcal{L}_\chi A_\mu + \partial_\mu\Lambda\,,\\
 \delta  l_\mu &=& \mathcal{L}_\chi l_\mu\,, \\
\delta \hat e_a\,^\mu  &=& \mathcal{L}_\chi \hat e_a\,^\mu +\Omega \hat\epsilon_a\,^{b}\hat e_b\,^\mu\,, 
 \end{eqnarray}
 where $\mathcal{L}_\chi$ is the Lie derivative along the vector $\chi^\mu = \xi^a\hat e_a\,^\mu +\kappa l^\mu$. Given the symmetry transformations it is possible to construct the covariant tensors $F_{\mu\nu}=2\partial_{[\mu }A_{\nu]}$, $\hat{R}_{\mu\nu}=2\partial_{[\mu }\hat{\omega}_{\nu]}$, $T_{\mu\nu}=2\partial_{[\mu }l_{\nu]}$, which will be used as the building blocks of the effective field theory we will discuss below. We will collectively refer to them as $X_{\mu\nu}$.

In addition, we also have  the Riemann tensor built from the metric, $R^\mu\,_{\nu\rho\lambda} $.

\section{The effective action} \label{sec:Eff}

In a quantum Hall state the fermions are gapped by the magnetic field and can be integrated out. The resulting effective action will be a local functional of the background sources, that must be invariant under the local symmetry transformations \eqref{eq:localtrans} and discrete symmetries of the fermionic action, in particular $PT$. Local invariance can be made manifest using covariant terms to construct the effective action. Discrete symmetries are then used to determine which of these terms are allowed. Once these are identified, the strategy will be to organize them in a derivative expansion, assuming the fields are slowly varying respect to the characteristic length and time scales of the problem. 
Given this, one can write down all possible terms up to a required order in the expansion and include the combinations that respect the discrete symmetries. We present the technical details in the Supplementary Material.

The derivative expansion is organized according to the physics of Landau levels in the semi-Dirac semimetal. Consistent with the anisotropic scaling of the fermionic action, we assign different orders to transverse and longitudinal derivatives ${\cal P}_a^b\partial_b \sim O(\epsilon^2)$, $l^a\partial_a\sim O(\epsilon)$, where $\epsilon \ll 1$ is a parameter we use to characterize the order. In general, the terms we present can have contributions at different orders in the expansion, so the order we assign always refers to the contribution of {\em lowest} possible order.

The magnetic field determines the gap, therefore we assign $B\sim O(\epsilon^0)$. Moreover we expect the chemical potential to be finite and non-zero, $A_t\sim O(\epsilon^0)$. Finally, in analogy with the isotropic case, we assume that electric field is small compared to the magnetic field, $E_x\sim E_y \sim O(\epsilon)$. This implies the following order in the expansion for the vector potentials 
$
A_x \sim B y\sim O(\epsilon^{-1}), \ \ A_y\sim O(\epsilon^0).
$
The metric and vielbeins are always non-vanishing, even when the geometry is flat, so $E^a\sim e_a \sim O(\epsilon^0)$. This implies that the components of the transverse spin connection will be higher order than the electric fields $\hat \omega_\mu \sim O(\epsilon^2)$. 

Now we use the fact that there is a vector $l_\mu$ (with transverse projector ${\cal P}_\mu^{\ \nu}$). 

For each field strength $X_{\mu\nu}$ ($X=F,\hat{R},T$)  we can define two vectors and a scalar
\be\label{eq:defX}
X_\mu=l^\alpha X_{\alpha\mu},\ \ \tilde{X}_\mu={\cal P}_{\mu\alpha}\varepsilon^{\alpha\nu\lambda} X_{\nu\lambda},\ \ X_\perp=\varepsilon^{\mu\nu\lambda}l_\mu X_{\nu\lambda}.
\ee
Note that the norm of $F_\mu$ (or the dual version $\tilde{F}_\mu$) is $\sim B$, so the order in the derivative expansion is $O(\epsilon^0)$. With the norm we can define a $O(\epsilon^0)$ scalar $\cB_y^2=F_\mu F^\mu\sim l_y^2B^2$. Moreover, one can construct a second $O(\epsilon^0)$ scalar using $F_\perp$: $\cB_0^2  = F_\perp^2 \sim  l_0^2 B^2 $. We will refer to them collectively as $\cB_A$.

Armed with the above formalism we are in a position to write down the effective action. To the lowest order $O(\epsilon^0)$ there are just two possible terms in the action
\be\label{eq:L0}
{\cal L}_0= -\cE(\cB_A)+\frac{\nu}{4\pi} \varepsilon^{\mu\nu\lambda}A_\mu\partial_\nu A_\lambda.
\ee
The first term is an arbitrary function of $\cB_A$. This can be thought of as energy density of the Quantum Hall State. The second term is the topological Chern-Simons term.
At the next level in the expansion ($O(\epsilon)$) the terms that respect $PT$ symmetry are
\be
{\cal L}_1=\sum_{i=1}^6 c_i^{PT}(\cB_A) S_i^{PT}
\ee
where
\be
\begin{split}
&
S^{PT} =\{S_1, \ S_2, \ F_\perp S_3,\  F_\perp S_4, \ S_5 ,\ S_6\}, \\ 
&c^{PT}(\cB_A)=\{c_1, \ c_2,\ \bar{c}_3,\  \bar{c}_4,\ c_5 ,\ c_6\}  ,
\end{split}
\ee
and we define 
\be
\begin{split}
&S_1= l^\mu\partial_\mu \cB_y, \ \ S_2= l^\mu\partial_\mu \cB_0,\ \ S_3=  F^\mu T_\mu, \ \  S_4=  \tilde{F}^\mu T_\mu, \\ 
&S_5=\varepsilon^{\mu\nu\lambda}F_\mu \partial_\nu F_\lambda,\ \ S_6=\varepsilon^{\mu\nu\lambda}F_\mu \partial_\nu \tilde{F}_\lambda.
\end{split}
\ee
The anisotropy leads to a plethora of new terms at $O(\epsilon^2)$, which we list in the Supplemental Material. However, given our interest in topological transport we note that out of all these terms the one associated the so-called Euler current $\sim A_\mu J_E^\mu(u)$ is of particular importance because it gives rise to topological transport \cite{Golkar:2014wwa,Golkar:2014paa}.  The current is defined using the Riemann tensor and unit norm vectors $u^\mu$, $u^\mu u_\mu=\sigma$, where $\sigma=\pm 1$
\be
J_E^\mu(u)=\frac{1}{8\pi}\varepsilon^{\mu\nu\lambda}\varepsilon^{\alpha\beta\gamma}u_\alpha\left( \nabla_\nu u_\beta \nabla_\lambda u_\gamma+\frac{\sigma}{2}R_{\nu\lambda\beta\gamma}\right).
\ee
From all the possible terms there are only three independent corresponding to $u^\mu\propto l^\mu, F^\mu, \widetilde{F}^\mu$, and only the last one contributes to $PT$ preserving topological transport with a coefficient $\kappa$. Another $O(\epsilon^2)$ term that will be relevant is  $ T_\perp$, which is of the form of a torsional Hall viscosity involving only the longitudinal projection of the vielbeins\footnote{Even though the torsion field $T^a\,_{\mu\nu}=  -2(\partial_{[\mu}E\,^a_{\nu]}-\omega ^{a}_{\; \; b [\nu}E^b\,_{\mu]})$ vanishes by construction, the anisotropy allows one to introduce the Lorentz invariant one form $l_\mu=l_aE^a\,_\mu$ with field strength $T_{\mu \nu} .  $}. An analogous term was discussed in \cite{Copetti:2019rfp}.

Having an effective action we can now proceed to study responses of an anisotropic Quantum Hall state to external perturbations.

\section{Hall transport}  \label{sec:Hall}
We apply the formalism developed so far to study DC responses to electric field, strain and vorticity. Among these responses are topological responses whose importance stems from the fact that they often provide us with universal characteristics of quantum states. Such universal features are indispensable for a better understanding of strongly coupled fractional Hall states. 

We define the transverse and longitudinal components of the stress tensor from the variation of the effective action with respect to vielbeins and gauge fields

\be
\delta S= -\int  d^3 x \; \sqrt{-g}  \; (\tau _\mu\,^a \delta e_a\,^\mu+j^\mu \delta A_\mu )
\ee
The simplest of these responses is the Hall conductivity. It follows from the Chern-Simons term and it is given by
\begin{align}
     \sigma_{x y }   =  \frac{\nu}{2 \pi },
\end{align}
which is the same as in the isotropic case. We will next focus on the response of the stress tensor to applied strains. In our geometric formulation it will correspond to the metric fluctuations. Such fluctuations can be easily embedded in our construction as independent variations of vielbeins. A second variation of the action with respect to vielbeins gives the stress-stress two-point functions from which the transport coefficients can be extracted via Kubo formulas for the viscosity tensor through appropriate projections \cite{bradlyn2012read}
\be
\eta^{\mu \ \nu}_{\ a \ b}=\lim_{\omega\to 0}\frac{i}{\omega}\vev{\tau^\mu _{\ a}\tau^\nu_{\ b}}.
\ee
Below we show that anisotropic states preserving $PT$ symmetry can have two independent Hall contributions to the viscosity tensor corresponding to such responses.
 The vielbein variations naturally include non-isotropic and non-symmetric viscous responses. Therefore our formulation is analogous to related frameworks that introduce a non-zero torsion \cite{Hughes2011,Hughes2013,Copetti:2019rfp,Valle2015,Jensen:2014aia,Tutschku:2020drw,deBoer:2020xlc}. Before extracting the coefficients we note that the time-reversal odd viscosity tensor in a $PT$-invariant theory can be decomposed as follows
\be
 \eta^ {ijkl}= 8\eta^{\text{iso}} P_{\text{iso}}^{ijkl} +8\eta^{\text{nem}} P_{\text{nem}}^{ijkl}+8\eta^{\text{vor}} P_{\text{vor}}^{ijkl}  ,
\ee
where the projection tensors are defined as
\begin{eqnarray}
P_{\text{iso}}^{ijkl} & =& -\frac{1}{16}\left(\epsilon^{ik}\delta^{jl}+(i\leftrightarrow j)+(k\leftrightarrow l)+(ik \leftrightarrow jl) \right),\nonumber \\
P_{\text{nem}}^{ijkl} &= & \frac{1}{8} l^{(i} \tilde{l}^{j)} \delta^{lk}-(ij \leftrightarrow kl),  \\
P_{\text{vor}}^{ijkl} &= & \frac{1}{8}l^{[i} \tilde{l}^{j]}\delta^{lk}-(ij \leftrightarrow kl),\nonumber
 \end{eqnarray}
 where $\tilde{l}^i = \epsilon^{ij } l_j$. Using the above formulae we can extract Hall responses in the semi-Dirac phase (see the Supplement for details). The first coefficient is a modification of the isotropic contribution
\be
	\label{vis1}
	\eta^{\text{iso}}        = \frac{\kappa}{4\pi}B- \frac{1}{2} c_5 B^2-\frac{1}{2} f_{11}.
\ee
In addition there is another Hall viscosity encoding responses of symmetric strains, present due to the anisotropy. These properties are characteristic for the nematic phase so we refer to it as nematic \cite{You2014}. 
\be
	\label{vis2}
	\eta^{\text{nem}}        = \frac{1}{2}   c_5 B^2+ \frac{1}{2} f_{11}.
\ee
We do not find any other anisotropic contribution to the symmetric part of viscosity tensor in the semi-Dirac phase, although arguments based on a continuous symmetry group do not forbid a third coefficient. Finally we identify, hitherto neglected, a response to vorticity that appears in the semi-Dirac phase
\be
	\label{vis3}
	\eta^{\text{vor}}        =  \frac{1}{2} c_5 B^2-\frac{1}{2}f_{11},
\ee 
however, notice that this term is also responsible for a response in the anti-symmetric stress tensor due to symmetric strain.

In general, the coefficients are arbitrary functions of $B$ field and $\Delta$. For $\Delta \ll   B$ the dependence will be fixed by the scaling symmetry of the semi-Dirac point, up to corrections suppressed by $\Delta/B$. The Chern-Simons and Euler current terms have dimensionless coefficients, while 
\be
c_5\sim B^{-1},\ \ f_{11} \sim B.
\ee 	
Therefore, at the semi-Dirac point both odd viscosities have a similar dependence with the magnetic field $\eta_T\sim \eta_I\sim B$, but deviations are expected for $\Delta\neq 0$.

\section{Discussion}  \label{sec:dis}

Recent interest in odd responses is stimulated by experiments with fluids that break parity. Such fluids emerge in the context of electron hydrodynamics as well as biological set-ups that can be modelled by fluid membranes. Anisotropy can naturally appear for such flows and could modify flow solutions and physical outcome. 

Motivated by Quantum Hall physics of semi-Dirac critical phases we have developed a formalism that allows one to write down effective actions for such systems. A key ingredient in the construction is a covariant prescription allowing a coupling to geometry. The anisotropy is generated by a spatial vector that we treat as an independent field. Equipped with this construction we proposed an expansion scheme that allowed us to write down an effective action for the semi-Dirac phase. As an application of the formalism we calculated topological responses for this system showing independent contributions to the viscosity tensor. We noted that $PT$ symmetry restricts the components to two independent coefficients. Finally, we showed that semi-Dirac materials are a natural hosts for torsional-like responses.

In general all the coefficients in the effective action (except the filling fraction and $\kappa$) can vary continuously with the ratio $|\Delta/B^{\frac{2}{3}}|$, as long as both quantities are nonzero. For $|\Delta/B^{\frac{2}{3}}| \ll 1$ we expect a universal dependence of the coefficients with the magnetic field, with powers that are fixed by the scaling symmetry of the semi-Dirac phase. In the opposite limit one abandons the regime of the semi-Dirac phase, so the derivative expansion we have introduced is not expected to remain valid.

\section{Acknowledgements}
We thank Christian Copetti and Karl Landsteiner for discussions. R. L., F. P.-B. and P. S. acknowledge financial support by the Deutsche Forschungsgemeinschaft (DFG, German Research Foundation) under Germany's Excellence Strategy through W\"{u}rzburg-Dresden Cluster of Excellence on Complexity and Topology in Quantum Matter - ct.qmat (EXC 2147, project-id 390858490). C.H.~has been partially supported by the Spanish grant PGC2018-096894-B-100 and by the Principado de Asturias through the grant GRUPIN-IDI/2018 /000174.

\bibliographystyle{apsrev4-1}

\bibliography{AnisotropicHall}

\end{document}


\title{Supplementary Material for ``Quantum Hall effective action for anisotropic Dirac semi-metal"}
\author{Carlos Hoyos}
\affiliation{Department  of  Physics  and  Instituto  de  Ciencias  y  Tecnolog\'{i}as  Espaciales  de  Asturias  (ICTEA)\\
Universidad  de  Oviedo,  c/  Federico  Garc\'{i}a  Lorca  18,  ES-33007  Oviedo,  Spain}
 \author{Ruben Lier}
\author{Francisco Pe\~{n}a-Benitez}
\author{Piotr Sur\'owka}

 \affiliation{Max-Planck-Institut  f\"ur Physik komplexer Systeme, N\"othnitzer Str. 38, 01187 Dresden, Germany}

\begin{abstract}
\end{abstract}


\maketitle

\section{Discrete symmetries}
With our conventions, the parity, time reversal and charge conjugation transformations of relativistic fermions are \cite{Deser1982}
\be
P^{-1} \psi(t,x,y) P= \sigma_2 \psi(t,-x,y),\ \
T^{-1} \psi(t,x,y) T= \sigma_1\sigma_3 \psi(-t,x,y),\ \
C^{-1} \psi(t,x,y) C=\sigma_1 \psi^*(t,x,y)
\ee
In addition, we also consider a $\pi$ rotation on the $(x,y)$ plane,  that we identify with $P_*$ \cite{Gusynin:2007ix}
\be
P_*^{-1} \psi(t,x,y) P_*= \sigma^3\psi(t,-x,-y)
\ee
The Dirac-like kinetic term is invariant and the mass-like kinetic term is $P$ and $T$ odd, so the coupling $m$ is $P$ and $T$ odd. Then we can assign the following parities to the couplings in the fermionic action
$$
\begin{array}{c|cccc}
  & C & P & T  & P_* \\ \hline
m & + & - & - & + \\
l^a & + & + & + & -
\end{array}
$$


The transformation of the background fields is collected in the table below
$$
\begin{array}{c|cccc}
  & C & P & T  & P_* \\ \hline
\partial_\mu & + & (-1)^{\delta_{\mu,1}} & (-1)^{\delta_{\mu,0}} & -(-1)^{\delta_{\mu,0}} \\
A_\mu  & - & (-1)^{\delta_{\mu,1}} & -(-1)^{\delta_{\mu,0}} &  -(-1)^{\delta_{\mu,0}} \\
e_a^{\ \mu} & + & (-1)^{\delta_{a,1}+\delta_{\mu,1}} & (-1)^{\delta_{a,0}+\delta_{\mu,0}} &   (-1)^{\delta_{a,0}+\delta_{\mu,0}} \\
g_{\mu\nu} & + & (-1)^{\delta_{\mu,1}+\delta_{\nu,1}} & (-1)^{\delta_{\mu,0}+\delta_{\nu,0}}  &   (-1)^{\delta_{\mu,0}+\delta_{\nu,0}} \\
\hat \omega_\mu & + & -(-1)^{\delta_{\mu,1}} & -(-1)^{\delta_{\mu,0}} &   -(-1)^{\delta_{\mu,0}}\\ 
l^\mu & + & (-1)^{\delta_{\mu,1}} & (-1)^{\delta_{\mu,0}} & -(-1)^{\delta_{\mu,0}} \\
F_{\mu\nu} & - & (-1)^{\delta_{\mu,1}+\delta_{\nu,1}} & -(-1)^{\delta_{\mu,0}+\delta_{\nu,0}} & (-1)^{\delta_{\mu,0}+\delta_{\nu,0}}  \\
\hat R_{\mu\nu} & + & -(-1)^{\delta_{\mu,1}+\delta_{\nu,1}}  & -(-1)^{\delta_{\mu,0}+\delta_{\nu,0}} &  (-1)^{\delta_{\mu,0}+\delta_{\nu,0}} \\
T_{\mu\nu} & + &   (-1)^{\delta_{\mu,1}+\delta_{\nu,1}} &  (-1)^{\delta_{\mu,0}+\delta_{\nu,0}} & (-1)^{\delta_{\mu,0}+\delta_{\nu,0}}
\end{array}
$$

\section{Derivative expansion}

In the identification of possible terms appearing in the effective action we will use covariant expressions. 
The microscopic theory respects $PT$ symmetry, so all the allowed terms should be $PT$ even. In the covariant construction we present those terms are automatically symmetric under charge conjugation (as defined for relativistic $2+1$ fermions) and the $P_*$ symmetry. Among the $PT$ even terms, the most relevant for us will be those that are $P$ odd and $T$ odd. 

To leading order $O(\epsilon^0)$, the possible terms are given by
\be
{\cal L}_0=-{\cal E}(\cB_a)+\frac{\nu}{4\pi}\varepsilon^{\mu\nu\lambda}A_\mu \partial_\nu A_\lambda.
\ee
The Chern-Simons term is $P$ and $T$ odd and the energy density is even. 

Before imposing $PT$ symmetry, at $O(\epsilon)$ the possible terms are
\be
{\cal L}_1=\sum_{i=1}^6 c_i(\cB_a) S_i+\sum_{i=1}^6 \overline{c}_i(\cB_a) F_\perp S_i,
\ee
where
\be
\begin{split}
&S_1= l^\mu\partial_\mu \cB_y, \ \ S_2= l^\mu\partial_\mu \cB_0,\ \ S_3=  F^\mu T_\mu, \ \  S_4=  \tilde{F}^\mu T_\mu, \\ 
&S_5=\varepsilon^{\mu\nu\lambda}F_\mu \partial_\nu F_\lambda,\ \ S_6=\varepsilon^{\mu\nu\lambda}F_\mu \partial_\nu \tilde{F}_\lambda.
\end{split}
\ee
The $PT$ symmetric terms are
\be
S_1, \ S_2, \ F_\perp S_3,\  F_\perp S_4, \ S_5 ,\ S_6.
\ee
From these, the $P$ and $T$ odd terms are $F_\perp S_3$ and $S_5$.

At $O(\epsilon^2)$ we can add Euler current terms, that give potential contributions to the Hall viscosity \cite{Golkar:2014wwa,Golkar:2014paa}
\be
\sim A_\mu J_E^\mu(u).
\ee
The current is defined with  the Riemann tensor and unit norm vectors $u^\mu$, $u^\mu u_\mu=\sigma$, where $\sigma=\pm 1$
\be
J_E^\mu(u)=\frac{1}{8\pi}\varepsilon^{\mu\nu\lambda}\varepsilon^{\alpha\beta\gamma}u_\alpha\left( \nabla_\nu u_\beta \nabla_\lambda u_\gamma+\frac{\sigma}{2}R_{\nu\lambda\beta\gamma}\right).
\ee
Although we have several $O(\epsilon^0)$ vectors at our disposal, it turns out that there are only two independent $PT$ even conserved Euler currents, for $u^\mu=v^\mu$ and $u^\mu=j^\mu$, where
\be
v^\mu=\frac{\tilde{F}^\mu}{\sqrt{-\tilde{F}_\mu \tilde{F}^\mu}},\ \ j^\mu=\frac{F^\mu}{\sqrt{F_\mu F^\mu}}.
\ee
The norms are $v_\mu v^\mu=-1$ and $j^\mu j_\mu=+1$. A similar term would be a Wen-Zee term mixing the gauge field with the transverse spin connection $\sim A\wedge \hat{\omega}$, but it is $PT$ odd and thus excluded. Then we are left with
\be\label{eq:euler}
{\cal L}_{\text{Euler}}=\kappa A_\mu J^\mu_E(v)+\kappa' A_\mu J_E^\mu(j).
\ee
Of the two possible terms, the one with coefficient $\kappa$ is $P$ and $T$ odd, while the one with coefficient $\kappa'$ is $P$ and $T$ even.

In addition to the Euler current terms, we list here other possible $O(\epsilon^2)$ contributions. Before imposing $PT$ symmetry the possible terms are
\be
\begin{split}
&{\cal L}_2=\sum_{i<j} d_{ij}(\cB_a) S_i S_j+\sum_{i=1}^6 e_i(\cB_a) l^\sigma \partial_\sigma S_i\\
&+\sum_{i<j} \overline{d}_{ij}(\cB_a) F_\perp S_i S_j+\sum_{i=1}^6 \overline{e}_i(\cB_a) F_\perp l^\sigma \partial_\sigma S_i\\
&+\sum_{i=1}^{11} f_i(\cB_a) S_i^{(2)}+\sum_{i=1}^{11}\overline{f}_i(\cB_a) F_\perp S_i^{(2)}.
\end{split}
\ee
Where
\be
\begin{split}
&S_1^{(2)}=F^\mu \partial_\mu \cB_y ,\ \ S_2^{(2)}=F^\mu \partial_\mu \cB_0 ,\\ 
&S_3^{(2)}=\tilde{F}^\mu \partial_\mu \cB_y ,\ \ S_4^{(2)}=\tilde{F}^\mu \partial_\mu \cB_0 , \\ 
& S_5^{(2)}=\nabla_\mu F^\mu ,\ \ S_6^{(2)}=\nabla_\mu \tilde{F}^\mu \ \ S_7^{(2)}=F^\mu \partial_\mu F_\perp,\\ 
&S_8^{(2)}= T_\mu T^{\mu} ,\ \ S_9^{(2)}=\varepsilon^{\mu\nu\lambda} T_\mu \partial_\nu F_\lambda ,\\ 
&S_{10}^{(2)}=\varepsilon^{\mu\nu\lambda}\tilde{T}_\mu \partial_\nu F_\lambda, \ \ S_{11}^{(2)}=T_\perp .
\end{split}
\ee
There are too many terms to list them all. Restricting to $P$ and $T$ odd terms we have, for terms of type $S_i^{(2)}$:
\be
S_7^{(2)}, S_{11}^{(2)}, \ \ F_\perp \times \{ S_1^{(2)}, S_2^{(2)},S_5^{(2)},S_{10}^{(2)}\}
\ee
The terms of the form $l^\sigma \partial_\sigma S_i$ are
\be
l^\sigma\partial_\sigma S_5, \ \ l^\sigma\partial_\sigma  S_7,\ \  F_\perp l^\sigma\partial_\sigma S_3.
\ee
The quadratic terms $S_i S_j$ are
\be
\begin{split}
&\{S_1,S_2\}\times \{ S_5,S_7\},\ \ S_3 S_4,\ \ S_5 S_6,\ \ S_6 S_7, \\
&\{S_1,S_2\} \times F_\perp S_3,\ \ F_\perp S_3 S_6,\ \ F_\perp S_4 S_5,\ \ F_\perp S_4 S_7.
\end{split}
\ee

\subsection{Simplifications}

If we restrict to homogeneous DC transport of the spatial components of the current and the stress tensor, then we can neglect all the terms that have spatial derivatives or more than one time derivative, and reduce to a geometry that is non-trivial only in the spatial directions. With these simplifications $l^\mu$ will be zero along the time direction, so terms with longitudinal derivatives $\sim l^\sigma\partial_\sigma$ can be set to zero. $F_\perp$ will be proportional to the electric fields and all the terms with $F_\perp$ factors can be dropped. Since all the $O(\epsilon)$ terms involve at least one derivative, all the $O(\epsilon^2)$ terms that are quadratic or have a derivative of the $O(\epsilon)$ terms can be dropped as well.

The leftover potential $P$ and $T$ odd contributions are the Chern-Simons term for the gauge field, the $O(\epsilon)$ terms $S_5$ and $S_6$ and, at $O(\epsilon^2)$ the Euler current \footnote{The Euler current has two derivatives, but after integrating by parts the gauge potential produces a $\sim B$ factor times a single-derivative term}  with coefficient $\kappa$ and the term $S_{11}^{(2)}$. However, since the spatial components of $\tilde{F}_\mu$ are proportional to the electric fields, $S_6$ can be discarded as well. This leaves, in addition to the Euler current and Chern-Simons, the two contributions $S_5$  and $S_{11}^{(2)}$

\section{Stress correlators from odd terms}
In this section we show how to use the Kubo formalism to extract the coefficients corresponding to viscous transport for our effective action. The second variation with respect to the vielbein determines the correlator
\be
\vev{\tau^\mu_{\ a} \tau^\nu_{\ b}}=\frac{\delta^2 {\cal S}}{\delta E^b\,_{\nu}\delta E^a\,_{\mu}}.
\ee
We will also assume a background magnetic field
\be
F_{\mu\nu}=B \epsilon_{\mu\nu 0}.
\ee
For plane wave metric perturbations $\sim e^{-i \omega t}$ one gets:
\be
\begin{split}  
                 &\vev{\tau^i_{\ a}\tau^j_{\ b}}_{\kappa }=  \frac{i \omega}{8 \pi }   \kappa  B \left(\epsilon^{ij}\delta^{kl}+(i\leftrightarrow k)+(j\leftrightarrow l)+(ij \leftrightarrow kl) \right) \delta_{a  k}  \delta_{b   l }   \\ 
&\vev{\tau^i_{\ a}\tau^j_{\ b}}_{f_{11}}=  - 2i \omega f_{11}l_a l_b \epsilon^{ij}, \\
&\vev{\tau^i_{\ a}\tau^j_{\ b}}_{c_5}=  -  2i \omega c_5 B^2 \epsilon_{ab} l^i l^j  .
\end{split}
\ee
Here the vielbeins are flat, i.e. $l_a = \delta^y_a$ . The labels of the correlators correspond to the first term in \eqref{eq:euler}, $S_{11}^{(2)}$ and $S_{5}$ respectively. 

\be
\eta^{i\ j}_{\ a \ b}=\lim_{\omega\to 0}\frac{i}{\omega}\vev{\tau^i_{\ a}\tau^j_{\ b}},
\ee
we find that the effective action terms produce the following three viscosity tensors:
\be
\begin{split}
   &(\eta^{i\ j}_{\ a \ b})_{\kappa}=  -  \frac{1}{8 \pi }    \kappa  B\left(\epsilon^{ij}\delta^{kl}+(i\leftrightarrow k)+(j\leftrightarrow l)+(ij \leftrightarrow kl) \right)  \delta_{a  k }  \delta_{b   l }  ,    \\ 
&(\eta^{i\ j}_{\ a \ b})_{f_{11}}=  2 f_{11}l_a l_b \epsilon^{ij}, \\
&(\eta^{i\ j}_{\ a \ b})_{c_5}=   2c_5 B^2 \epsilon_{ab} l^i l^j   .
\end{split}
\ee
Using the projectors
\begin{eqnarray}
P_{\text{iso}}^{ijkl} & =& -\frac{1}{16}\left(\epsilon^{ik}\delta^{jl}+(i\leftrightarrow j)+(k\leftrightarrow l)+(ik \leftrightarrow jl) \right),\nonumber \\
P_{\text{nem}}^{ijkl} &= & \frac{1}{8} l^{(i} \tilde{l}^{j)} \delta^{lk}-(ij \leftrightarrow kl), \nonumber \\
P_{\text{vor}}^{ijkl} &= & \frac{1}{8} l^{[i} \tilde{l}^{j]}\delta^{lk}-(ij \leftrightarrow kl).
 \end{eqnarray}
With $\tilde{l}^i = \epsilon^{ij } l_j$. These projectors are normalized so that for all of them $(P_{ij k l })^2 = \frac{1}{8} $. One finds the following three independent contributions to viscous transport for our effective action:
\be
\begin{split}
&\eta_{\text{iso}}=   \frac{ \kappa  B }{4 \pi }    - \frac{1}{2}  f_{11}  - \frac{1}{2}  c_5 B^2 ,\\
&\eta_{\text{nem}}=  \frac{1}{2}   f_{11}   +  \frac{1}{2}  c_5 B^2  , \\
&\eta_{\text{vor}}=  - \frac{1}{2}   f_{11}  + \frac{1}{2}   c_5 B^2 .
\end{split}
\ee

Here we used:
\be
\eta^{(a)}  = \eta^{i\ j}_{\ a \ b} P^{(a)}_{ikjl} \delta^{a k} \delta^{b l } 
\ee

\bibliographystyle{apsrev4-1}

\bibliography{AnisotropicHall}